\documentclass[journal=jpcbfk,manuscript=article]{achemso}
\usepackage{longtable}
\usepackage{multirow}
\usepackage{amsmath}
\usepackage{subfigure}
\usepackage[usenames,dvipsnames]{color}
\usepackage{soul}
\usepackage{bm}
\newcommand{\norm}[1]{\left\lVert#1\right\rVert}
\usepackage{xr}

\usepackage{multicol}
\usepackage{wrapfig}
\usepackage{enumitem}
\usepackage{bm}
\usepackage{outlines} %for itemize with subitems
\SectionNumbersOn

\author{Silvan K\"aser} \affiliation[University of Basel]{Department
  of Chemistry, University of Basel, Klingelbergstrasse 80 , CH-4056
  Basel, Switzerland.}

\author{Debasish Koner} \affiliation[University of Basel]{Department
  of Chemistry, University of Basel, Klingelbergstrasse 80 , CH-4056
  Basel, Switzerland.}

\author{Anders S. Christensen} \affiliation{Institute of Physical
  Chemistry and National Center for Computational Design and Discovery
  of Novel Materials (MARVEL), Department of Chemistry, University of
  Basel, Klingelbergstrasse 80, CH-4056 Basel, Switzerland}

\author{O. Anatole von Lilienfeld} \affiliation{Institute of Physical
  Chemistry and National Center for Computational Design and Discovery
  of Novel Materials (MARVEL), Department of Chemistry, University of
  Basel, Klingelbergstrasse 80, CH-4056 Basel, Switzerland}
\email{anatole.vonlilienfeld@unibas.ch}

\author{Markus Meuwly} \affiliation[University of Basel]{Department of
  Chemistry, University of Basel, Klingelbergstrasse 80 , CH-4056
  Basel, Switzerland.}  \email{m.meuwly@unibas.ch}

\title{ML Models of Vibrating H$_2$CO: Comparing Reproducing Kernels,
  FCHL and PhysNet}
\begin{document}

\date{\today}

\begin{abstract} 
Machine Learning (ML) has become a promising tool for improving the
quality of atomistic simulations. Using formaldehyde as a benchmark
system for intramolecular interactions, a comparative assessment of ML
models based on state-of-the-art variants of deep neural networks
(NN), reproducing kernel Hilbert space (RKHS+F), and kernel ridge
regression (KRR) is presented. Learning curves for energies and atomic
forces indicate rapid convergence towards excellent predictions for
B3LYP, MP2, and CCSD(T)-F12 reference results for modestly sized (in
the hundreds) training sets. Typically, learning curve off-sets decay
as one goes from NN (PhysNet) to RKHS+F to KRR (FCHL). Conversely, the
predictive power for extrapolation of energies towards new geometries
increases in the same order with RKHS+F and FCHL performing almost
equally. For harmonic vibrational frequencies, the picture is less
clear, with PhysNet and FCHL yielding respectively flat learning at
$\sim 1$ and $\sim 0.2$ cm$^{-1}$ no matter which reference method,
while RKHS+F models level off for B3LYP, and exhibit continued
improvements for MP2 and CCSD(T)-F12. Finite-temperature molecular
dynamics (MD) simulations with the same initial conditions yield
indistinguishable infrared spectra with good performance compared with
experiment except for the high-frequency modes involving hydrogen
stretch motion which is a known limitation of MD for vibrational
spectroscopy. For sufficiently large training set sizes all three
models can detect insufficient convergence (``noise'') of the
reference electronic structure calculations in that the learning
curves level off. Transfer learning (TL) from B3LYP to CCSD(T)-F12
with PhysNet indicates that additional improvements in data efficiency
can be achieved.
\end{abstract}

\section{Introduction}
\begin{figure}[h]
\centering
\includegraphics[width=1\textwidth]{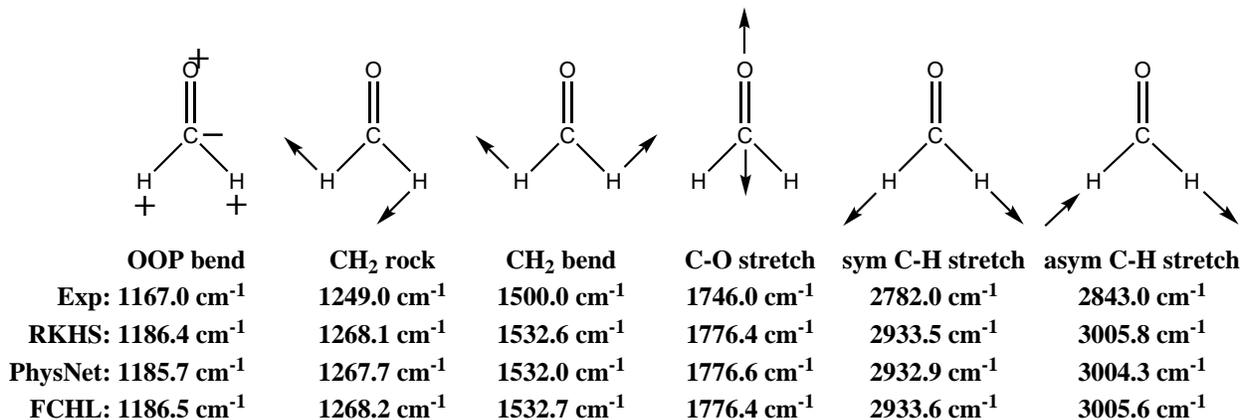}
\caption{Molecular structure and the directions of the normal modes
  for formaldehyde (H$_2$CO). ``+'' and ``-'' indicate motions with
  respect to the plane containing all four atoms. Experimental IR
  frequencies ($\nu_4$, $\nu_6$, $\nu_3$, $\nu_2$, $\nu_1$, and
  $\nu_5$) in ascending order of the
  experiments\cite{herndon2005determination} are reported together
  with harmonic frequencies obtained from ML models trained on the
  largest CCSD(T)-F12 data set size.}
\label{fig:formaldehyde}
\end{figure}

With the advent of machine learning (ML) in the physical sciences a
paradigm shift has taken place
\cite{behler2016perspective,von2018quantum,
  butler2018machine,MM.rev.jcp:2020}. In particular for molecular
sciences where the interaction between particles is of central
importance for developing quantitatively meaningful models, ML offers
many opportunities for improved and computationally efficient modeling
of systems. This also leads to the question which - if any - of the
existing and currently pursued approaches to represent inter- and
intramolecular potential energy surfaces is most advantageous. Such an
assessment includes questions pertaining to how ``data hungry'' a
particular approach is (i.e. how much data is required to achieve a
given level of accuracy for a particular property), how accurate the
resulting PES is, whether the model can be used to extrapolate to
unknown regions not sampled by the reference data and finally, whether
computed observables from the models differ or whether they are
largely insensitive to the representation and its quality given the
same reference data set. All these points will be assessed in the
present work for formaldehyde (H$_2$CO, see
Figure~\ref{fig:formaldehyde}).\\

\noindent
Formaldehyde is a small molecule for which very
high-level calculations have already been presented
\cite{zhang2004global, wang2017new, karton2006w4} and 
experimental reference data is available to compare with
\cite{herndon2005determination}. Apart from its suitability for
in-depth theoretical study, formaldehyde is also interesting because
it (i) is an important precursor in chemical industries
\cite{reuss2000formaldehyde} (ii) plays an ubiquitous role in many
domains including biology, atmosphere, toxicology, interstellar
chemistry \cite{european2014endogenous, wang2017sources,
  zhang2018formaldehyde, snyder1969microwave} and (iii) was first
implicated in the phenomenon of 'roaming'
\cite{townsend2004roaming}.\\

\noindent
Earlier theoretical work on formaldehyde includes a global PES based
on CCSD(T)/aug-cc-pVTZ and MR-CI/aug-cc-pVTZ calculations for which
different fits are smoothly joined using switching
functions\cite{zhang2004global} and a newer, refined global PES
employing multi reference configuration interaction (MRCI/cc-pVTZ)
calculations.\cite{wang2017new} The root mean squared error (RMSE) of the fit
to the CCSD(T)/aug-cc-pVTZ data ranged from 277 cm$^{-1}$ to 648
cm$^{-1}$ (0.8 to 1.9 kcal/mol), depending on the energy range
considered (10000 to 38500 cm$^{-1}$).\cite{zhang2004global} For this
fit, Morse-type variables had been used. For the more recent global
PES,\cite{wang2017new} fit to permutationally invariant polynomials
and based on MRCI reference data, the averaged RMS error was 100
cm$^{-1}$. Both surfaces were used to study roaming.\\

\noindent
The present work compares three currently available ML approaches
using the same reference data sets computed at three representative
levels of quantum chemical rigor (Hybrid density functional
approximation (B3LYP)\cite{becke1993becke, lee1988development},
M{\o}ller Plesset 2$^{\rm nd}$ order perturbation theory
(MP2)\cite{moller1934note}, and Coupled Cluster Single Doubles
perturbative Triples (CCSD(T)-F12))\cite{adler2007simple}.  The three
ML methods are also meant to be representative in that they range from
a purely kernel-based approach (reproducing kernel Hilbert space -
RKHS\cite{rabitz:1996, MM.rkhs:2017} plus forces
(RKHS+F)\cite{kon20:rkhsf}) to a purely neural network (NN) based
approach (PhysNet~\cite{MM.physnet:2019}), and include the FCHL
representation~\cite{fchl:2018} within kernel ridge regression (KRR).
It should be mentioned that there is obviously a considerably larger
number of alternative methods, equally suited quantum chemistry and ML
methods that could have been used just as well. However, the limited
present selection is largely due to the focus on the particular
system, formaldehyde, and its particular properties relevant to
intramolecular interactions.\\

\noindent
The present work is structured as follows. First, the three ML methods
are introduced, followed by a description of how the data sets were
generated and how vibrational spectra were computed. The results
compare the mutual performance of the ML methods by considering energy
and force learning curves, harmonic frequencies and IR spectra from
finite-temperature MD simulations at the highest level of quantum
chemical theory. This is followed by a discussion and conclusion.\\

\section{Theory}
In the following the machine learning methods, the generation of the
data sets including the structure sampling procedure and the quantum
chemical calculations are explained. Then, the computation
of vibrational spectra is reviewed.

\subsection{Machine Learning Methods}
In this section the ML methods used in the present work are introduced
and specifics about them are given.

\subsubsection{PhysNet} High-dimensional PESs can be constructed using
PhysNet which is a NN of the ``message-passing''
type.\cite{gilmer2017neural} So-called feature vectors are learned to
encode a representation of the local chemical environment of each
atom. In an iterative fashion, the initial feature vector depending
only on nuclear charges $Z_i$ and Cartesian coordinates $\bm{r}_i$ of
all atoms $i$ is adjusted by passing ``messages''
between atoms. Based on the learned feature vectors, PhysNet predicts
atomic energy contributions and partial charges for arbitrary
geometries of the molecule. The total potential energy of the system
corresponds to $E = \sum_{i=1}^N E_i$, where $E_i$ are the atomic
energy contributions. The partial charges $q_i$ are corrected to
assure that the total charge of the system is conserved according to
the following scheme:
\begin{align}
 \tilde{q}_i = q_i - \frac{1}{N}\left(\sum_{j=1}^N q_j - Q\right)
\end{align}
Here, $\tilde{q}_i$ are the corrected partial charges, $q_i$ are the
partial charges predicted by PhysNet and $Q$ is the total charge of the
system \cite{MM.physnet:2019}. The forces $\bm{F}_i$ required to run
MD simulations are calculated analytically by reverse mode automatic
differentiation \cite{baydin2017automatic}.\\

\noindent
During training the PhysNet parameters are adjusted to best describe
the reference energies, forces and dipole moments from quantum
chemical calculations. The optimization uses ``adaptive moment
estimation'' (ADAM) \cite{kingma2014adam}. For a detailed description
of the PhysNet architecture as well as fitting procedure the reader is
referred to Reference~\citenum{MM.physnet:2019}. 

\subsubsection{Kernel-based methods} 
Next, two kernel-based methods to represent potential energy surfaces
are detailed. The first method (``RKHS+F'') is based on reproducing
kernel Hilbert spaces\cite{aronszajn:1950} and uses a distance-based
representation, while the second method uses a regressor from Gaussian
process regression combined with the
Faber-Christensen-Huang-Lilienfeld representation (``FCHL''), which is
a refined spatial representation based on radial and angular
spectra.\cite{fchl:2018,christensen2020} Kernel-based methods explore
the possibility to formulate the task of fitting a PES as an inversion
problem. The theory of these methods asserts that for $N$ given data
points $\mathbf{x}_i$ of a function $f_i = f(\mathbf{x}_i)$, the value
of $f(\mathbf{x})$ at an arbitrary point $\mathbf{x}$ can always be
approximated as a linear combination of kernel
products\cite{scholkopf2001generalized}
\begin{equation}
f(\mathbf{x}) = \sum_{i = 1}^{N} \alpha_i K(\mathbf{x},\mathbf{x}_i)
\label{eq:RKHS_function}
\end{equation}
Here, the $\alpha_i$ are regression coefficients and
$K(\mathbf{x},\mathbf{x'})$ is a kernel function. The coefficients
$\alpha_i$ can be determined from inverting
 \begin{equation}
 f_j = \sum_{i = 1}^{N} \alpha_i K_{ij}
 \label{eq:RKHS_coefficient_relation}
 \end{equation}
using, e.g.\ Cholesky decomposition\cite{golub2012matrix}, where
$K_{ij} = K(\mathbf{x}_i,\mathbf{x}_j)$ is the symmetric,
positive-definite kernel matrix. With this, the value of
$f(\mathbf{x})$ for an arbitrary argument $\mathbf{x}$ can be
calculated using Eq.~\ref{eq:RKHS_function}.\\

\noindent
{\bf Specifics for RKHS+F}\\ For higher-dimensional problems the RKHS+F
method used here constructs $D$-dimensional kernels as tensor products
of one-dimensional kernels $k(x,x')$
\begin{equation}
K(\mathbf{x},\mathbf{x_i}) = \prod_{d=1}^{D} k^{(d)}(x^{(d)},x_i^{(d)})
\label{eq:multidimensional_kernel}
\end{equation}
For the kernel functions $k(x,x')$ it is possible to encode physical
knowledge, such as their long range interactions which has been done
for weakly interacting systems.\cite{hutson.rkhs:00} The general
expression for the 1D kernel function used here is
 \begin{equation}
 k^{[n,m]} = n^2x_{>}^{-(m+1)}B(m+1,n)
 _2F_1\left(-n+1,m+1;n+m+1;\frac{x_<}{x_>}\right )
\label{rprk}
\end{equation}
where, $n$ and $m$ are the smoothness and asymptotic reciprocal power
parameters, whereas $x_<$ and $x_>$ are the smaller and larger value
of $x$, respectively. $B(a,b)$ in Eq. \ref{rprk} is the beta function
$B(a,b) = \frac{(a-1)!(b-1)!}{(a+b-1)!}$ and $_2F_1(a,b;c;z)$ is
Gauss' hypergeometric function.\cite{rabitz:1996} For different types of
bonds it may be necessary to choose different kernel functions.\\

\noindent
Within a many body expansion, the total potential energy of a system
can be decomposed into a sum of $p$-body interactions $V^{(p)}$. For a
molecule with $n$ atoms, each $p$-body term consists of $^nC_p$
$p$-body interactions, where $^nC_p$ is the binomial coefficient. The
total potential for an $n$-atomic species is therefore
 \begin{equation}
  V = \sum_{p=1}^n \sum_{i=1}^{^nC_p} V_i^{(p)}
\label{mbe}
 \end{equation}
In practice Eq. \ref{mbe} is truncated at $p=3$ or 4. Here, $p=4$ was
used, i.e. all many-body terms were included.\\
 
 \noindent
In the present study, each term of the $p$-body interaction energy is
represented as an $M$-dimensional ($M = $ $^pC_2$) reproducing kernel
constructed from $M$ reciprocal power kernels for $M$ interatomic
distances $r_j$. The full kernel is then
\begin{equation}
K({\bf r}, {\bf r'}) = \sum_{p=1}^n
\sum_{l=1}^{^nC_p}\prod_{j=1}^{^pC_2} k_j(r_j,r'_j)
\label{fullkernel1}
 \end{equation}
and
  \begin{equation}
     V({\bf r}) = \sum_{i=1}^N \alpha_i K({\bf r}, {\bf r'})
     \label{fullkernel2}
 \end{equation}
Here, ${\bf r}$ is a vector containing all pairwise interatomic
distances of an $n$-atomic system, ${\bf r} = \{r_h|h=1,2,3
\cdots,^nC_2$\}. In the present study, $k^{[3,5]}$, $k^{[3,1]}$ and
$k^{[3,0]}$ kernel functions are used to construct
mono/multidimensional kernels for 2-, 3-, and 4-body interaction
energies, respectively.\\

\noindent
The symmetry of a molecule is explicitly included in the total kernel
polynomial $K({\bf r}, {\bf r'})$ (see Eq. \ref{fullkernel1}) by
expanding it as a linear combination of all equivalent structures of a
molecule. Examples are shown in Ref. \citenum{kon20:rkhsf} for CH$_4$
and CH$_2$O.\\
 
\noindent
{\bf Specifics for FCHL}\\ An alternative approach to representing the
molecules comes from recent developments in machine learning
representations for molecules.\cite{felixgoogle2017} These often allow
for improved learning rates at the cost of increased model
complexity. The FCHL representation\cite{fchl:2018} used in this work
describes the atomic environment of an atom as histograms based on the
radial distribution of surrounding atoms and Fourier terms for the
angular distributions. This makes it possible to train models that
span molecules and materials of varying sizes and chemical
composition. For forces and energies, the ``FCHL19'' representation is
used, which is a coarse-grained, discretized, and numerically
efficient implementation of FCHL with pre-optimized hyper-parameters
for force and energy learning.\cite{christensen2020}\\

\noindent
FCHL relies on the ``localized'' kernel \textit{ansatz},
to ensure size-extensivity and permutational atom index invariance.\cite{bpkc2010} 
Here, it is used together with a Gaussian kernel
function where the kernel element between
two molecules corresponds to the sum of pair-wise Gaussian kernel functions between
atoms in the respective two molecules:
\begin{equation}
  \label{eq:kernel_screening}
     \mathbf{K}_{ij} = \sum_{I \in i} \sum_{J \in i} \delta_{Z_I
       Z_{J}} \exp\left(-\frac{\| \mathbf{x}_I - \mathbf{x}_{J}
       \|^2_2}{2\sigma^2}\right)
\end{equation}
where $\mathbf{x}_I$ and $\mathbf{x}_{J}$ are the representation of
the $I$'th and $J$'th atoms in the molecules $i$ and $j$,
respectively, and the Kronecker-$\delta$ between $Z_I$ and $Z_J$
(their atomic numbers) ensuring bagging, as demonstrated previously to
be advantageous for universal quantum ML models based on the
bag-of-bonds representation~\cite{BOB}.\\

\noindent
{\it Regression for RKHS+F:} Derivatives of the potential with respect
to the distance coordinates can be calculated analytically up to order
$(n-1)$ by simply replacing the kernel polynomial $K({\bf r}, {\bf
  r'})$ by their derivatives $K'({\bf r}, {\bf r'})$. Using the chain
rule, gradients with respect to Cartesian coordinates can also be
obtained, which are also available from the electronic structure
calculations. For an RKHS+F-based representation of the PES, the set
of linear equations can be written as a matrix equation
\begin{equation}
  \label{rkhs+f}
    \begin{bmatrix}
        \mathbf{v} \\ \mathbf{f}
    \end{bmatrix} = \begin{bmatrix}
       \mathbf{K} & \\
       -\frac{\partial}{\partial \mathbf{r}}\mathbf{K} &
    \end{bmatrix} \bm{\alpha},
\end{equation}
where $\mathbf{v}$ and $\mathbf{f}$ are vectors containing energies
and forces, respectively, and $\bm{\alpha}$ is a vector containing a
set of regression coefficients.  For an $n-$atomic species the matrix
in the left hand side becomes rectangular with dimension $(3n+1)N
\times N$ Eq. \ref{rkhs+f} are solved using a least square fitting
algorithm. The `DGELSS' subroutine in the LAPACK library is used to
solve the set of linear equations.\\

\noindent
{\it Regression for FCHL:} For the model using the FCHL
representation, a model for energies and forces is implemented
similarly to what is commonly known from Gaussian process regression
and kernel-ridge regression with
derivatives.\cite{GAPtutorial,RasmussenWilliams} Here, a PES can be
regressed from the training set of molecules with reference energy and
force labels. By placing the kernel functions and corresponding kernel
derivatives on the molecules in the training set, the set of equations
to train or predict energies and forces
are\cite{GAPtutorial,sonjamathias}
\begin{equation}
  \label{eq:gpr_derivative}
    \begin{bmatrix}
        \mathbf{v} \\ \mathbf{f}
    \end{bmatrix} = \begin{bmatrix}
       \mathbf{K} && -\frac{\partial}{\partial \mathbf{r}^\top}\mathbf{K} \\
       -\frac{\partial}{\partial \mathbf{r}}\mathbf{K} && \frac{\partial^2}{\partial \mathbf{r}\partial \mathbf{r}^\top}\mathbf{K} 
    \end{bmatrix} \bm{\alpha}.
\end{equation}
This is akin (except for using the Hessian) to Eq. \ref{rkhs+f} for
the RKHS+F approach, but with an extended set of basis functions.
Similarly to both RKHS+F and PhysNet, forces can be evaluated as the
derivative of the energy, which is crucial for energy conservation.\\

\noindent
The optimal regression coefficients $\bm{\alpha}$ can then be
obtained, for example, by minimizing the following cost function:
\begin{equation} \label{eq:gpr_cost}
J(\bm{\alpha}) = \frac{1}{2}\norm{ \begin{bmatrix}
       \mathbf{K} && -\frac{\partial}{\partial \mathbf{r}^\top}\mathbf{K} \\
       -\frac{\partial}{\partial \mathbf{r}}\mathbf{K} && \frac{\partial^2}{\partial \mathbf{r}\partial \mathbf{r}^\top}\mathbf{K} 
    \end{bmatrix}  \bm{\alpha} -  \begin{bmatrix}
        \mathbf{v}^\mathrm{ref} \\
        \mathbf{f}^\mathrm{ref} 
    \end{bmatrix}  }_2^2 
    + \frac{\lambda}{2} \bm{\alpha}^\top  \begin{bmatrix}
       \mathbf{K} && -\frac{\partial}{\partial \mathbf{r}^\top}\mathbf{K} \\
       -\frac{\partial}{\partial \mathbf{r}}\mathbf{K} && \frac{\partial^2}{\partial \mathbf{r}\partial \mathbf{r}^\top}\mathbf{K} 
    \end{bmatrix}  \bm{\alpha} 
\end{equation}
where $\lambda$ is a regularizer that puts a small penalty on large
regression coefficients and ensures numerical stability to the
minimizer. Compared with RKHS+F, FCHL uses second derivatives (see Eq. \ref{eq:gpr_derivative}) which
increases computational cost but also further improves performance\cite{christensen2020}
(\textit{vide infra}).\\

\noindent
For dipole moments, the analytical implementation of the FCHL L2-norm
in Eqn.~\ref{eq:kernel_screening} (i.e.~``FCHL18'') is used as
previously described.\cite{christensen2018operators} The
implementation adds a dependence on an externally applied field
$\mathbf{E}$ to the representation via a set of fictitious atomic
partial charges. For non-zero fields, this component is crucial
in order to ensure that the uniqueness condition of a representation can be met~\cite{FourierDescriptor}.
The result is a physics based representation
for machine learning models of the dipole moment, as obtained by differentiating
ML model of the energy with respect to field. More specifically, the following relations
of the kernel-based energy model are used:
\begin{equation}
    \mathbf{v} = \mathbf{K}   \bm{\alpha}
\end{equation}
and the relationship between energy and dipole moment
\begin{equation}
  \bm{\mu} = -\frac{\partial}{\partial \Vec{\mathbf{E}}}\mathbf{v}
  % = -\frac{\partial}{\partial \Vec{\mathbf{E}}}\mathbf{K}   \bm{\alpha}
\end{equation}
leads to
\begin{equation}
        \bm{\mu} =\left[ -\frac{\partial}{\partial \Vec{\mathbf{E}}}\mathbf{K} \right] \bm{\alpha}
\end{equation}
The regression coefficients can be obtained by minimizing a cost
function such as
\begin{equation} \label{eq:gdml_lagrangian}
J(\bm{\alpha}) = \tfrac{1}{2}\norm{  
    \left[ -\tfrac{\partial}{ \partial \Vec{\mathbf{E}}}\mathbf{K} \right]
          \bm{\alpha} - \bm{\mu}^\mathrm{ref} }_2^2 
\end{equation}
 in a \textit{least squares} fit. In practice, a singular-value
 decomposition of the kernel derivative matrix is used via the LAPACK
 subroutine DGELSD.\\

\noindent
For simplicity the Gasteiger-Marsili charge
model\cite{GASTEIGER19803219} as implemented in Open
Babel\cite{OBoyle2011a} is used to obtain the fictitious charges but
the learned model has been found to depend little on the choice of the
fictitious charges\cite{christensen2018operators}, as long as they are
physically reasonable, since the numerical values of the partial
charges will be absorbed into the regression coefficients.

\section{Methods}
\subsection{Quantum Chemical Calculations}
The reference energies, forces and dipole moments are obtained from
quantum chemical calculations at different levels of theory and using
different quantum chemical programs. They include the B3LYP\cite{becke1993becke, lee1988development}/cc-pVDZ\cite{dunning1989gaussian}
level of theory calculated using Orca\cite{neese2012orca}, and the
MP2\cite{moller1934note}/aug-cc-pVTZ\cite{kendall1992electron}
and the CCSD(T)-F12\cite{adler2007simple}/aug-cc-pVTZ-F12\cite{peterson2008systematically}
levels of theory obtained from Molpro \cite{MOLPRO} calculations. Loose convergence
criteria on the Hartree-Fock reference wave function can lead to noise
in the energies and forces used in the training. Therefore, tighter
convergence criteria were used as follows. For Orca the SCF
convergence criterion (``VeryTightSCF'') and the DFT integration grid
(``Grid7'' and ``NoFinalGrid'') were used. For calculations with
Molpro the convergence criteria are tightened using the ``gthresh''
keyword and set to ``gthresh,orbital=1.d-8, gthresh,energy=1.d-11''
and ``gthresh,orbital=1.d-8, gthresh,energy=1.d-12'' for the
MP2/aug-cc-pVTZ and the CCSD(T)-F12/aug-cc-pVTZ-F12 calculations,
respectively.\\

\subsection{Data Set}
To assess the performance of the different approaches, two data sets
were generated. Set1 contains $N_{\rm tot} = 4001$ H$_2$CO structures
including the optimized H$_2$CO structure. It was randomly split into
subsets (training and test set) of different sizes. The geometries
were generated by means of normal mode sampling.\cite{smith2017ani}
Starting from the optimized H$_2$CO structure at the B3LYP/cc-pVDZ
level of theory and knowing the normal mode coordinates together with
their harmonic force constants, distorted conformations were obtained
by randomly displacing the atoms along the normal modes.  To capture
the equilibrium, room temperature and higher energy regions of the PES
the normal mode sampling was carried out at eight different
temperatures (10~K, 50~K, 100~K, 300~K, 500~K, 1000~K, 1500~K,
2000~K). For each temperature 500 structures were generated.\\

\noindent
For assessing the extrapolation capabilities of the different ML
methods, a second data set (Set2) was generated which also contains
distorted structures, not sampled in Set1. For this, 2500 structures
were generated from normal mode sampling carried out at 5000 K.\\

\noindent
The ML models are trained on different training set sizes from Set1,
including $N_{\rm train} = 100$, 200, 400, 800, 1600, and 3200
structures and tested on the remaining $N_{\rm tot} - N_{\rm train}$
structures. Therefore, the indices of the structures are shuffled
(based on a seed) and the training set was taken to be $ [0:N_{\rm
    train}]$. For each training set size (and ML method and level of
theory) a total of five independent models are trained, where a
different seed is used for the shuffling of the indices. To guarantee
direct comparability of the different ML methods the models are
trained and tested on exactly the same reference data.\\

\subsection{Vibrational Spectra}
Vibrational spectra were computed by means of normal mode analysis and
finite-temperature molecular dynamics (MD) simulations. These
simulations were carried out using the atomic simulation environment
(ASE)\cite{larsen2017atomic} together with the best respective ML
model trained on the CCSD(T)-F12 data. For direct comparison of the
different ML models the MD simulations were started from identical
initial conditions, i.e. the same molecular geometry and initial
momenta. The initial geometry was the optimized H$_2$CO structure from
the FCHL-based model which is identical to the minimized structures
from the other two methods, see Tables~S2 to
S4 and the momenta were drawn randomly from a
Maxwell-Boltzmann distribution and scaled to correspond to exactly
300~K. The optimized geometries from the ab initio calculations as
well as optimized using the ML models are listed in
Tables~S1 to S4.\\

\noindent
First, the molecule was equilibrated in the $NVE$ ensemble for 50~ps,
followed by MD simulations with a time step of $\Delta t = 0.5$~fs for
a total of 200~ps. When running the simulation with PhysNet, the
molecular dipole moment $ \bm{\mu} = \sum_{i=1}^N q_i\bm{r}_i$ is
calculated and saved simultaneously for each snapshot of the
trajectory, whereas for FCHL computing the dipole moment is a post
processing step. With RKHS+F no dipole moment was learned. Instead,
PhysNet was used to obtain the molecular dipole for the structures
sampled in the simulations with the RKHS+F PES.\\

\noindent
The infrared spectra are then obtained from the Fourier transform of
the dipole-dipole auto-correlation function $C(t) = \langle
\bm{\mu}(0)\bm{\mu}(t) \rangle$. Using the efficient Fast Fourier
Transform algorithm from the numpy python library, the transform
$C(\omega)$ is obtained using a Blackman filter.  In addition, for
PhysNet 1000 independent trajectories were run following the same
protocol outlined above using the unscaled momenta drawn from a
Boltzmann distribution at 300 K. From this, a conformationally
averaged IR spectrum was calculated to test convergence.

\section{Results}
The performance of the models is determined by considering the mean
absolute error (MAE) and the RMSE for energy
$E$ and forces $F$. This is done for the test set and for the
extrapolation data set, the learning curves, for the harmonic
frequencies and the (anharmonic) frequencies obtained from finite
temperature MD. \\

\subsection{Learning Curves}
Learning curves report the out-of-sample prediction error as a
function of training set size. They are a useful way to compare
different ML techniques on the same footing and to assess how rapidly
they reach a particular accuracy.\\

\begin{figure}[ht]
\centering
\includegraphics[width=1\textwidth]{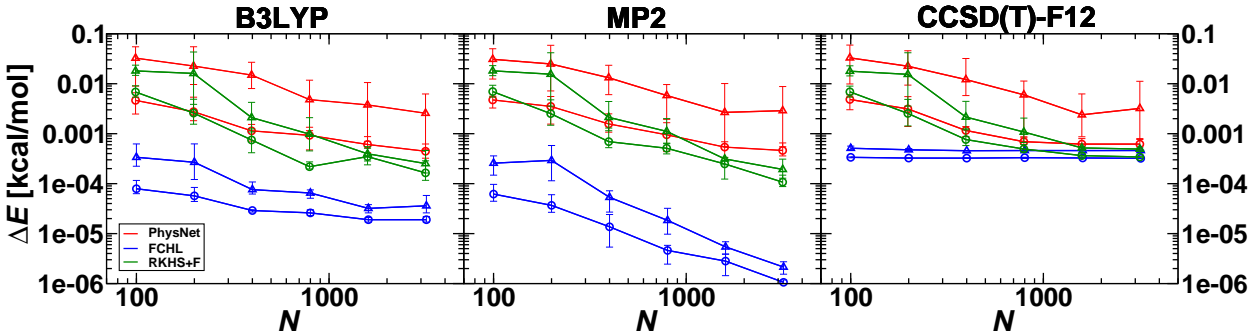}
\caption{Energy learning curves (Log-log plot) for the PhysNet- (red),
  FCHL- (blue) and RKHS+F- (green) based models. All models are
  trained on the same reference data with different data set sizes
  (100, 200, 400, 800, 1600, 3200) and on data calculated at different
  levels of theory (B3LYP, MP2, and CCSD(T)-F12 from left to
  right). The MAE is shown as a circle and the RMSE is shown as a
  triangle. $\Delta E$ corresponds to the energy error and the error
  bars indicate the minimum and maximum error. Every data point is an
  average over 5 models trained independently on the same data set
  size, but different samples from the full data set.}
\label{fig:energy_learning_curve}
\end{figure}

\noindent
Figure \ref{fig:energy_learning_curve} reports energy learning curves
of PhysNet, RKHS+F, and FCHL using the three quantum chemical reference
methods B3LYP, MP2, and CCSD(T)-F12.  For all reference methods
systematic improvement of predictive power is observed as the training
set size increases, reaching very good accuracies of at least
$10^{-3}$ kcal/mol (compared with $\sim 1$ kcal/mol from earlier work
fit to CCSD(T) reference data).\cite{zhang2004global} Among the ML
methods tested, FCHL yields the lowest errors, followed by RKHS+F and
PhysNet. For the two largest training set sizes (1600, 3200), the
learning curve of PhysNet ceases to learn, except for B3LYP. This
could be due to the fact that PhysNet, as it is common among ANNs,
represents a non-parametric supervised ML model.  Interestingly, the
deviation among MAE and RMSE is larger for PhysNet than for the
kernel-based methods. Significant differences among various error
measures of prediction error statistics of KRR machine learning models
were recently studied in great detail\cite{pernot:2020}, suggesting
that further analysis also including NN-based models is warranted. \\

\begin{figure}[h]
\centering \includegraphics[width=1\textwidth]{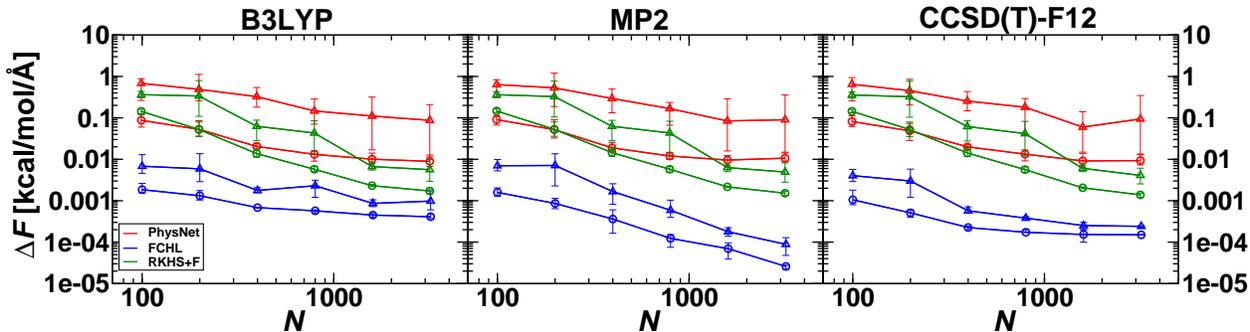}
\caption{Force learning curves (log-log plot) for the PhysNet- (red),
  FCHL- (blue) and RKHS+F- (green) based models. All models are
  trained with different data set sizes (100, 200, 400, 800, 1600,
  3200) and on data of different levels of theory. The MAE is shown as
  a circle and the RMSE is shown as a triangle. $\Delta F$ corresponds
  to the force error and the error bars indicate the minimum and
  maximum error. Every data point is an average over 5 models trained
  independently on the same data set size, but different samples from
  the full data set.}
\label{fig:force_learning_curve}
\end{figure}

\noindent
 The impact of reference method selection on learning curves is
 considerable: For MP2, all ML models exhibit steep learning curves
 which do not saturate, for the kernel based models in particular and
 with FCHL reaching a MAE of $10^{-6}$ kcal/mol.  By contrast, when
 using CCSD(T)-F12 as a reference, rapid convergence towards an error
 floor of several $10^{-4}$ kcal/mol is observed for all ML models.
 Learning curves for the B3LYP reference lie in between these two
 extremes, converging for FCHL towards an error floor of several
 $10^{-5}$~kcal/mol. The existence of such floors in learning curves
 of functional machine learning models suggests that there is
 ``noise'' in the data. Indeed, inspection of the literature indicates
 that the forces in MOLPRO at the CCSD(T)-F12 level are less accurate
 than machine-precision.\cite{werner:2018} Before using the data set
 in the current learning study, the existence of such noise-levels was
 not known to the authors. The learning curve of FCHL obtained for
 B3LYP references may display a similar effect, e.g.~resulting from
 noise due to unconverged integration settings. Additional testing
 supports this explanation: Using B3LYP calculations with standard
 convergence criteria for SCF iterations and integration grid, the
 learning curve floors are confirmed for all three ML models. It is,
 therefore, concluded that ML is capable of detecting such subtle
 convergence issues in unseen data sets. The force learning curves,
 see Figure \ref{fig:force_learning_curve}, and learning curves for
 the dipole moment (for PhysNet- and FCHL-based models, see
 Figure~S4) display a similar pattern
 as the energy learning curves. \\

\subsection{Extrapolation of the PESs}
The ability of ML-models to extrapolate to geometries outside the
interval covered by the reference data is particularly relevant for MD
simulations. Traditionally, energy functions (such as empirical force
fields\cite{mackerell2004} or variants
thereof\cite{kramer2012atomic,bereau2013}) are fit to parametrized
functions for which the short- and long-range part of the interaction
is given by the functional form. Such an approach allows one to tailor
in particular the long-range behaviour to represent the physically
known interactions.\cite{MM.heh2:2019} This is different for most
machine-learned PESs. As an exception, using RKHS+F provides the
possibility to choose a kernel that captures the leading long-range
part of the physical interaction.\cite{rabitz:1996,hutson.rkhs:00}\\

\begin{figure}[h!]
\centering
\includegraphics[width=1.0\textwidth]{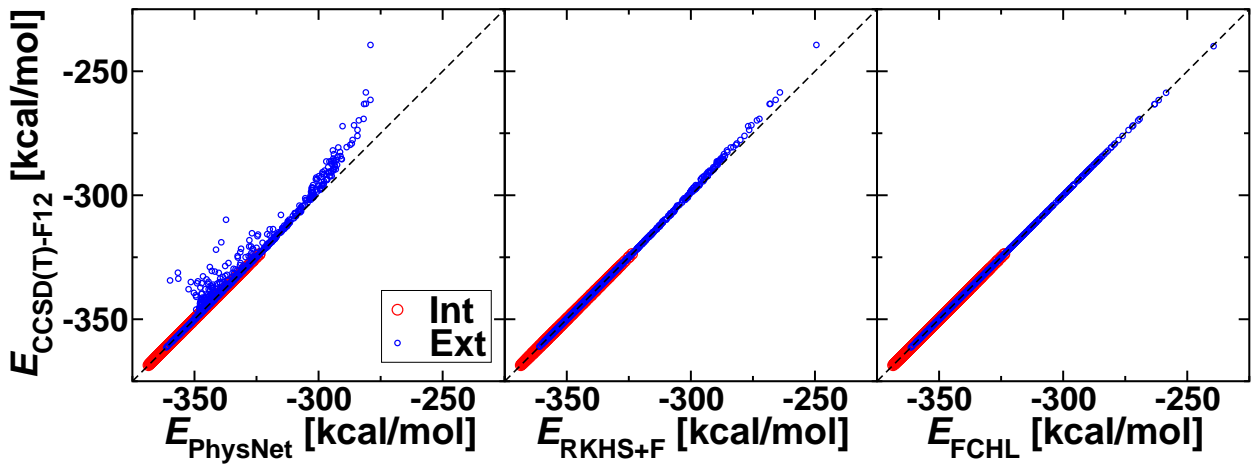}
\caption{Comparison of the reference ab initio CCSD(T)-F12 energies
  ($y-$axis) with the machine-learned energies ($x-$axis) for
  formaldehyde geometries in Set2 (2500 structures). Here the models
  trained on the largest data set size of 3200 and having the smallest
  MAE($E$) were used. The data points are shown in two distinct colors
  (and sizes for clarity) for the CO, CH, and OH bonds which cover
  intervals in the training data set bracketed by [1.14, 1.28], [0.88,
    1.40] and [1.83, 2.24] \AA\/, respectively. The red symbols are
  for structures in which {\it all} distances are inside the range
  sampled in the training set (1333 structures) whereas the blue
  symbols is for all other structures (1167 structures). PhysNet (left
  panel) performs well for unknown structures covered by the training
  set but fails for structures outside. RKHS+F (middle panel) is very
  reliable for all structures in Set2 except for those with the
  highest energy and FCHL (right panel) predicts all geometries
  reliably.}
\label{fig:corr_extrapol}
\end{figure}

\noindent
The extrapolation data set (Set2) sampled at higher temperature than
the training data set is examined and histograms for three bond
lengths (C--O, C--H and O--H) are shown in
Figures~S1 to S3 for both the
training and the extrapolation data set. It is apparent that
structures contained in Set2 reach more distorted geometries and were
never ``seen'' by the ML models. Figure~\ref{fig:corr_extrapol} shows
the comparison between the reference CCSD(T)-F12 and the
machine-learned energies from models trained on the largest reference
data set. The predictions from PhysNet either agree with the reference
(indicated by the red circles on the black dashed line) or yield lower
energies than the reference calculations. This is different for RKHS+F
and FCHL which reliably (blue symbols in Figure
\ref{fig:corr_extrapol}) extrapolate to energies a factor of $\sim 3$
higher than the energy range covered by the training set (red symbols
in Figure \ref{fig:corr_extrapol}). The performance of FCHL is even
better than that of RKHS+F. The MAEs and RMSEs of the three ML methods
trained on different data set sizes and tested on Set2 are illustrated
in Figure~S5.\\

\subsection{Vibrational Spectroscopy}
\begin{table}[h]
\begin{tabular}{c|c|c|c|c|c}
[cm$^{-1}$] & Exp & CCSD(T)-F12 &RKHS+F& PhysNet & FCHL \\
\hline 
$\nu_1$  & 2782.0 & 2933.8 & 2933.5 & 2932.9 & 2933.6 \\
$\nu_2$ & 1746.0 & 1776.4 & 1776.4 & 1776.6 & 1776.4 \\
$\nu_3$ & 1500.0 & 1532.7 & 1532.6 & 1532.0 & 1532.7 \\
$\nu_4$ & 1167.0 & 1186.5 & 1186.4 & 1185.7 & 1186.5 \\
$\nu_5$ & 2843.0 & 3005.8 & 3005.8 & 3004.3 & 3005.6 \\
$\nu_6$ & 1249.0 & 1268.2 & 1268.1 & 1267.7 & 1268.2 \\
\hline
RMSE    &  & 93.4 & 0.1 & 0.9 & 0.1 \\
\end{tabular}
\caption{Comparison of the normal mode
  frequencies obtained from the different ML models with their
  reference frequency (CCSD(T)-F12) and with those from
  experiment\cite{herndon2005determination}. The frequency
  calculations were performed with the model trained on the largest
  data set of 3200 structures and having the lowest MAE($E$).
  The RMSEs between the experiment and the \textit{ab initio}
  frequencies (CCSD(T)-F12) and between the \textit{ab initio} reference frequencies and the ML
  predictions (RKHS+F, PhysNet and FCHL) are given.}
\label{tab:freq1}
\end{table}

\noindent
Normal mode frequencies were determined at the CCSD(T)-F12 level of
theory, see Table~\ref{tab:freq1}. Using the trained models on the
largest reference data set ($N=3200$) the harmonic vibrations were
calculated and are found to be in very close agreement with those from
the quantum chemical calculations at the same level of theory. RMSEs of 0.14,
0.86 and 0.12~cm$^{-1}$ were found for RKHS+F, PhysNet and FCHL, respectively.\\

\begin{figure}[ht]
\centering
\includegraphics[width=1\textwidth]{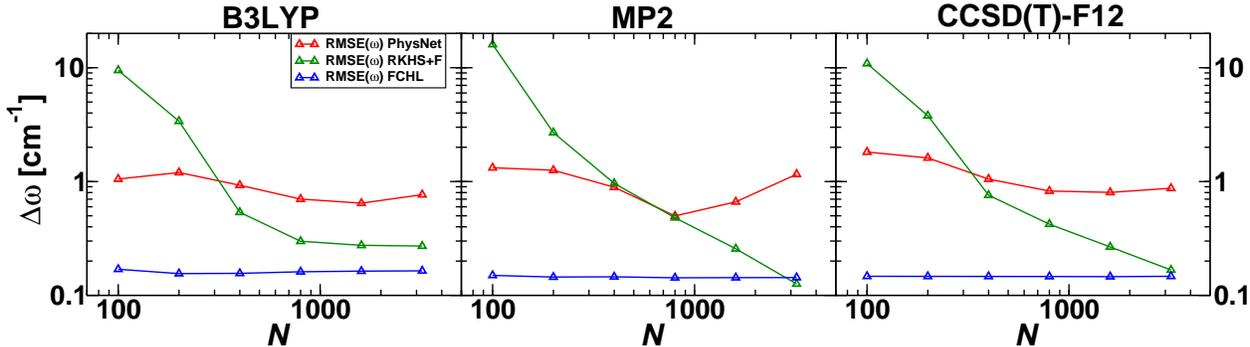}
\caption{Performance curves showing RMSE $\Delta \omega$ for harmonic
  frequencies between the reference {\it ab initio} values and
  respective prediction using the PhysNet- (red), RKHS+F- (green)  and
  FCHL- (blue) based models as a function of training set sizes.
  Left, mid, right panels correspond to training data from different
  levels of theory.}
\label{fig:lc_freq}
\end{figure}

\noindent
It is also of interest to determine harmonic frequencies from models
with smaller training sets in order to assess whether there is a
relationship between the accuracy of the machine-learned PES (see
Figure \ref{fig:energy_learning_curve}) and a particular observable,
here the normal mode frequency, see Figure~\ref{fig:lc_freq}. It is
found that for PhysNet the accuracy with which the reference quantum
chemical normal mode frequencies are predicted from the PhysNet-based
PES are uniformly within $\sim 1$ cm$^{-1}$, independent on the size
of the training set. Similarly, no data set size dependence is found
for the FCHL frequency predictions which accurately reproduce the
reference values with $\Delta\omega\sim 0.1$~cm$^{-1}$. The RKHS+F
based predictions, however, show a $\Delta\omega\sim 10$~cm$^{-1}$ for
the smallest training set size ($N=100$) and reach accuracies similar
to the FCHL models for the largest training set sizes. In other words,
all models are able to accurately predict the normal mode frequencies
of H$_2$CO at all three levels of theory considered here but the
number of training points $N$ to do may differ. \\

\noindent
Infrared spectra from MD simulations further probe the regions around
the minimum of the PES and provide an additional way to validate the
trained ML-models. It is also possible to directly compare these
spectra with experiments although for the high frequency modes it is
known that the computed band positions can be inaccurate due to
limited sampling of the anharmonicities because zero point vibrational
energy is not included in classical MD
simulations.\cite{qu2018ir,qu2018quantum,MM.oxa:2017}\\

\begin{figure}[H]
\centering
\includegraphics[width=0.9\textwidth]{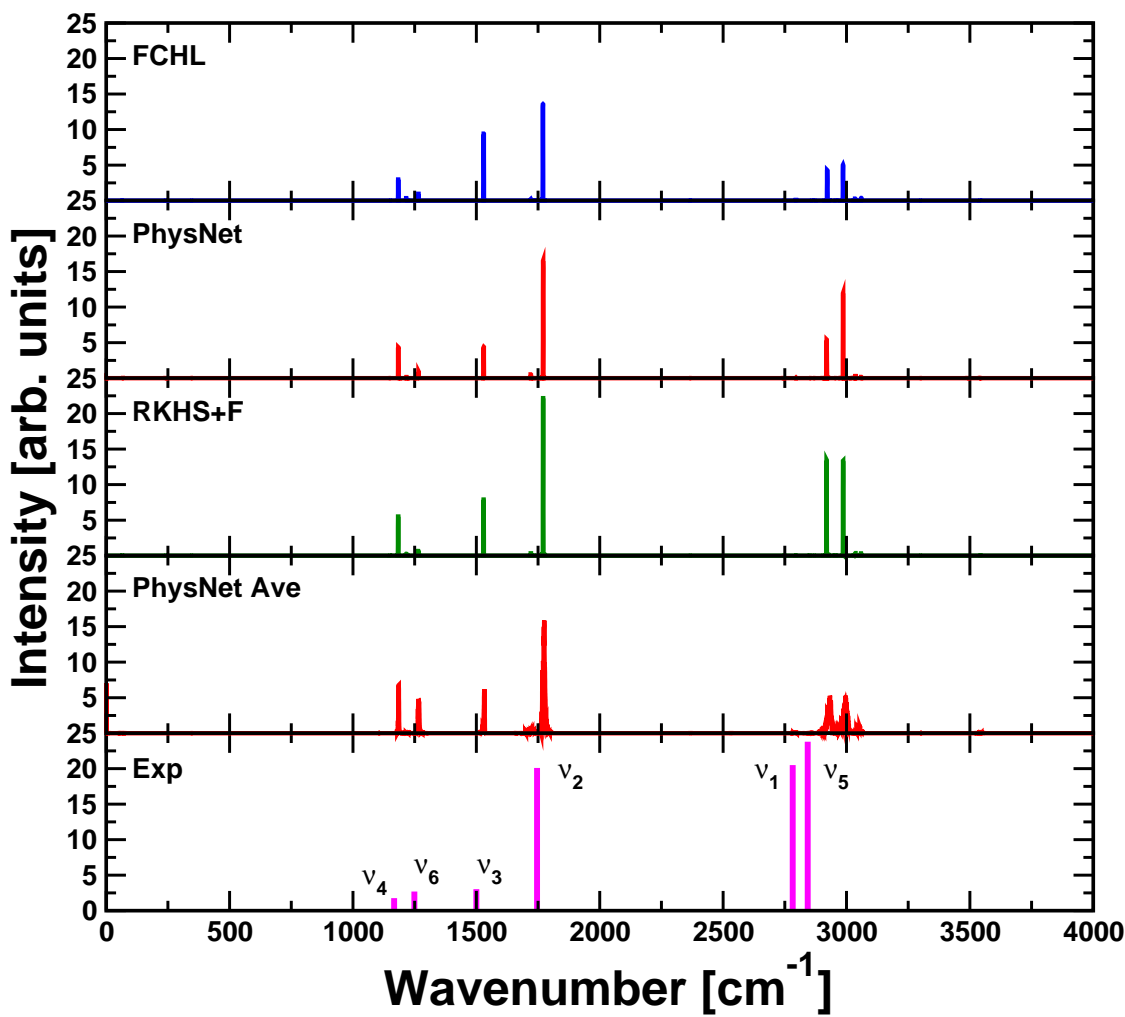}
\caption{ IR spectra from trajectories run on the different PESs. The
  trajectories from panels FCHL, PhysNet and RKHS+F were all started
  from the same initial conditions (geometry and momenta)
  corresponding to exactly 300~K. The RKHS+F method does not learn the
  molecular dipole moments and used PhysNet to predict the dipole
  moment of every snapshot along the RKHS+F trajectory. ``PhysNet
  Ave'' corresponds to the average over 1000 independent trajectories,
  each 200~ps in length. The trajectories are started from the same
  geometry but random momenta were drawn randomly from a
  Maxwell-Boltzmann distribution corresponding to 300~K. All
  simulations were run with the model trained on the largest data set
  of 3200 structures and having the lowest MAE($E$). The bottom
  panel reports the experimental infrared
  spectrum\cite{herndon2005determination} as a stick spectrum
  including intensities.}
\label{fig:infrared_comparison}
\end{figure}

\noindent
The IR spectra calculated from finite-$T$ MD simulations using the
FCHL-, PhysNet- and RKHS+F-based approaches (blue, red and green,
respectively) are shown in Figure~\ref{fig:infrared_comparison}. They
all agree very well with the experimentally determined band positions,
except for the CH stretching modes, and the relative intensities of
the bands are also qualitatively correct. For direct comparison, all
MD simulations were started from identical initial conditions. To
improve convergence of the spectra, an average over all 1000
independent runs using PhysNet is also reported. It was found that
even using random samples of 500 such spectra gives the same IR
spectrum. Therefore, the averaged spectrum shown can be considered
converged.\\

\section{Discussion}
The results indicate that all three ML-based approaches are successful
in correctly describing the near-equilibrium region of the PES. The
extrapolation capabilities of PhysNet are limited whereas RKHS+F and
FCHL provide robust global PESs with FCHL showing the best
performance, see Figure \ref{fig:corr_extrapol}. For energy and force
prediction and harmonic frequencies FCHL is best, followed by RKHS+F and
PhysNet, specifically for larger data sets. However, it should be
emphasized that the differences are generally small. For harmonic
frequencies none of the models differs by more than 1 cm$^{-1}$ from
the reference calculations, see Table \ref{tab:freq1}.\\

\begin{table}[h!]
\begin{tabular}{c|c|c|c}
\textbf{Mode} & \textbf{PhysNet} & \textbf{CCSD(T)-F12} &  $\Delta$    \\\hline
$\nu_1$ & 2933.62          & 2933.79          & 0.17 \\
$\nu_2$ & 1776.36          & 1776.39          & 0.03 \\
$\nu_3$ & 1532.56          & 1532.70           & 0.14 \\
$\nu_4$ & 1186.37          & 1186.46          & 0.09 \\
$\nu_5$ & 3005.68          & 3005.81          & 0.13\\
$\nu_6$ & 1268.21          & 1268.17          & 0.04
\end{tabular}
\caption{Harmonic frequencies in cm$^{-1}$ for a PhysNet model
  trained to convergence of the forces. All frequencies are within
  less than 0.2 cm$^{-1}$ of the reference CCSD(T)-F12 calculations,
  see also Table \ref{tab:freq1}.}
\label{tab:physnet_freq_force}
\end{table}

\noindent
For the PhysNet-based approach a few more tests have been carried
out. One of them concerns improving the harmonic frequencies ($\sim 1$
cm$^{-1}$ averaged difference compared with $\sim 0.1$ cm$^{-1}$ from
the two kernel-based methods) by applying a slightly different
learning protocol. For the PhysNet models presented above, training
was stopped upon convergence of the energy predictions. Although
further training would improve forces and dipole moments, the accuracy
in the energy prediction would deteriorate because forces and energies
are weighted differently in the loss function. Hence, an additional
PhysNet model was trained on 3200 data points to convergence of the
forces to investigate the accuracy of the harmonic frequencies. The
results, reported in Table~\ref{tab:physnet_freq_force}, are similar
to those from RKHS+F and FCHL with an averaged RMSE of $\Delta\omega
\approx 0.1$~cm$^{-1}$. On the other hand the MAE of the energy
(predicting the test set) increased by approximately one order of
magnitude to $\sim 4\cdot 10^{-3}$ ~kcal/mol, which is still very
accurate. Hence, for obtaining accurate harmonic frequencies a good
force-learned model can be advantageous.\\

\noindent
As the harmonic frequencies differ quite substantially from the
experimentally measured ones, it was also decided to compute the
anharmonic frequencies from PhysNet by supplying energies, forces and
the Hessian to the Gaussian09 quantum chemistry program. Because
direct comparison with the {\it ab initio} values was not possible for
the CCSD(T)-F12 level of theory, a comparison using the PhysNet MP2
model has been carried out and is reported in
Table~S5. The MP2 reference frequencies are
reproduced with an RMSE of $\sim1.0$ cm$^{-1}$ and the anharmonic
values with an RMSE of $\sim18.3$ cm$^{-1}$. Here, the largest
deviations are found for the high-frequency modes (deviation of $\sim
30$ cm$^{-1}$). Note that the MP2 model was trained only up to the
convergence of the energy and further training is expected to improve
anharmonic frequencies as well. Table \ref{tab:anh_freq} compares the
band centers from the finite-temperature IR spectra, the harmonic and
anharmonic frequencies from using PhysNet and the experimental
results. In particular for the stretch modes involving hydrogen atom
motion (CH symmetric and antisymmetric stretch) the improvement for
the anharmonic modes over harmonic frequencies and the MD simulations
is remarkable. But all other modes are also in considerably better
agreement with experiment with an RMSE of 10.8~cm$^{-1}$ between
experiment and anharmonic calculations. As PhysNet is the least
performing ML approach it is expected that similar calculations using
RKHS+F and FCHL trained on CCSD(T)-F12 reference data will be equally
good or even better.\\

\begin{table}[h]
\centering
\begin{tabular}{c|c|c|c|c}
\textbf{mode} & \textbf{IR/PhysNet} &\textbf{G09/PhysNet H} & \textbf{G09/PhysNet AH} & \textbf{Exp\cite{herndon2005determination}}\\\hline
$\nu_1$ & 2930.0 & 2932.9      & 2805.7        & 2782.0 \\
$\nu_2$ & 1773.0 & 1777.0      & 1741.5        & 1746.0 \\
$\nu_3$ & 1532.0 & 1532.8      & 1498.6        & 1500.0 \\
$\nu_4$ & 1185.0 & 1186.0      & 1170.9        & 1167.0 \\
$\nu_5$ & 2996.0 & 3004.1     & 2852.5        & 2843.0\\
$\nu_6$ & 1266.0 & 1268.3      & 1245.3        & 1249.0 \\
\hline
\textbf{RMSE} & 89.1 & 92.6 & 10.8 & 
\end{tabular}
\caption{Harmonic and anharmonic frequencies in cm$^{-1}$ for the
  optimized H$_2$CO structure compared with those from experiment
  ($\nu_i$)\cite{herndon2005determination}. The harmonic (G09/PhysNet
  H) and anharmonic (G09/PhysNet AH) frequencies are calculated
  with Gaussian 09 using the PhysNet PES trained on 3200 CCSD(T)-F12
  energies, forces and dipole moments. They are compared with IR
  center frequencies from experiment (Exp) and calculated from MD
  simulations (IR/PhysNet, see also Fig~\ref{fig:infrared_comparison},
  PhysNet Ave). The RMSE between experiment and computations is shown
  in the last row.}\label{tab:anh_freq}
\end{table}

\begin{figure}[h!]
\centering
\includegraphics[width=0.8\textwidth]{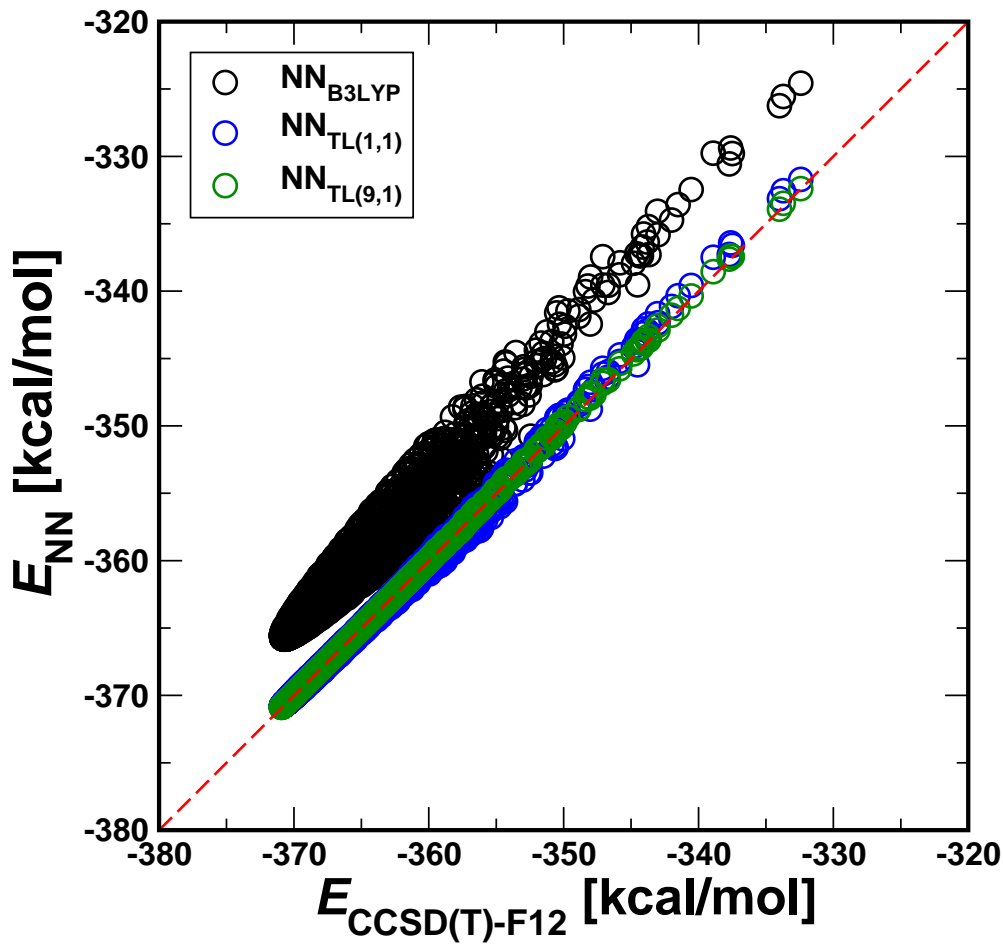}
\caption{Correlation between the energy predicted by different PhysNet
  models and the CCSD(T)-F12 energies for the same molecular
  geometries. The black circles show the performance of the B3LYP
  PhysNet model to predict the CCSD(T)-F12 energies. The blue and
  green circles show the performance of the B3LYP model transfer
  learned with $N_{\rm tot}^{\rm TL} = 2$ at the higher level of
  theory, respectively. Two structures suffice to eliminate the
  systematic shift whereas 10 structures where needed to reduce the
  scatter.}
\label{fig:corr_plot_TL}
\end{figure}

\noindent
The availability of PESs at different levels of theory also provides
the opportunity to discuss shortcuts to high level of theory PES
representations. Comparing B3LYP/cc-pVDZ energies (or predictions of
PhysNet trained on B3LYP data, PhysNet$_{\rm B3LYP}$) to
CCSD(T)-F12/aug-cc-pVTZ-F12 energies illustrates a systematic shift as
well as a regular scatter (see Fig.~\ref{fig:corr_plot_TL}, black
circles). Such correlations suggest that a combination of multiple
levels of theory during training will be beneficial. In the following,
``transfer learning'' (TL) was used\cite{taylor2009transfer,
  pan2009survey} although other methods, such as $\Delta$-Machine
Learning~\cite{DeltaPaper2015} (for kernel-based methods),
multi-fidelity learning~\cite{batra2019multifidelity}, or the
multi-level grid combination technique~\cite{zaspel2018boosting} could
also be used\\

\noindent
Starting from a PhysNet model trained at a lower level of theory
(B3LYP/cc-pVDZ), TL can be used to reach a higher level of theory
(CCSD(T)-F12/aug-cc-pVTZ-F12) at little additional cost. Thus, the
best B3LYP PhysNet model (as judged from the MAE($E$)) trained on 3200
H$_2$CO geometries is used as the reference and to initialize the
parameters of the TL model. Different TL models were generated based
on different training set sizes, with structures randomly chosen from
Set1 of the CCSD(T)-F12 data set. The following data set sizes $N_{\rm
  tot}^{\rm TL}(N_{\rm train}^{\rm TL},N_{\rm valid}^{\rm TL})$ were
considered for TL: 2(1,1), 10(9,1), 25(22,3), 50(45,5), 100(90,10),
and 200(180,20).\\

\noindent
The progress of TL PhysNet$_{\rm B3LYP}$ to CCSD(T)-F12 quality is
illustrated in Figure~\ref{fig:corr_plot_TL}. TL with $N_{\rm
  tot}^{\rm TL} = 2$ (blue circles) suffices to eliminate the
systematic shift between B3LYP and CCSD(T)-F12, whereas most of the
scattered data points are corrected with $N_{\rm tot}^{\rm TL} = 10$
(green circles).  Note that chemical accuracy (RMSE($E$) better than 1~kcal/mol
) is achieved with as little as two additional points at the
higher level of theory whereas a MAE($E{\rm )} = 0.004$~kcal/mol and a
RMSE($E$) = 0.006~kcal/mol is achieved with TL using $N_{\rm tot}^{\rm
  TL} = 200$. The performance based on MAEs and RMSEs of the remaining
data set sizes is summarized in Table~S6 and
illustrated as a learning curve in Figure~S6.\\

\noindent
Another measure for the performance of the TL models is the quality of
predicted normal mode frequencies compared with the CCSD(T)-F12
reference. For this the TL(180,20) model was used and the results are
summarized in Table~\ref{tab:freq}. This TL model reproduces the
reference CCSD(T)-F12 frequencies to within $\sim 5$ cm$^{-1}$ or
better. It is expected that a more careful selection of particular
geometries (e.g. geometries with displacements along normal modes)
will further improve the prediction with a smaller number of $N_{\rm
  tot}^{\rm TL}$ for accuracy as was recently shown for
malonaldehyde.\cite{mm.ht:2020}\\

\begin{table}[h]
\centering
\begin{tabular}{c|c|c||c}
 mode & TL(180,20) & PhysNet$_{\rm CCSD(T)-F12}$ &  CCSD(T)-F12 \\\hline
$\nu_1$ & 2930.8 & 2932.9 & 2933.79 \\
$\nu_2$ & 1777.0 & 1776.6 & 1776.39 \\
$\nu_3$ & 1529.9 & 1532.0 & 1532.70 \\
$\nu_4$ & 1179.1 & 1185.7 & 1186.46 \\
$\nu_5$ & 3000.9 & 3004.3 & 3005.81 \\
$\nu_6$ & 1266.3  & 1267.7 & 1268.17
\end{tabular}
\caption{Normal mode frequencies predicted by the transfer learned model
  TL(180,20) in comparison to a PhysNet trained on 3200 CCSD(T)-F12
  data points and with the ab initio CCSD(T)-F12 frequencies.}
\label{tab:freq}
\end{table}

\section{Conclusions}
We have investigated and compared the application of kernel and neural
network based ML models capable of generating fully dimensional PESs
for formaldehyde, and their application to vibrational spectroscopy.
Training/Test-set consistency runs indicate that the ML models achieve
such precision that noise levels even in unseen data can be
detected. We find it reassuring that all three ML models considered,
despite their differences in representation, functional form and
number of coefficients, result in overall excellent performance. In
particular, they are also applicable to extrapolation regimes, and
demonstrably useful for predicting experimental observables such as IR
spectra. With regards to the actual representation of atoms in
molecules, one can consider FCHL as intermediate between ``no
representation'' (RKHS+F) and a machine-learned representation
(PhysNet). Moreover, TL of PhysNet was demonstrated to result in substantial
improvements in data-efficiency. We expect our findings for machine
learning of high-quality PESs and harmonic frequency prediction to
also extend to larger molecules as has been recently demonstrated for
PhysNet\cite{mm.ht:2020,mm.atmos:2020} and molecules with up to 10
atoms using RKHS+F.\cite{kon20:rkhsf}\\

\section*{Supporting Information}
The supporting information reports the optimized structures of the
three learned models, histograms for bond lengths of Set1 and Set2,
additional learning curves and harmonic and anharmonic frequencies at
the MP2 level.

\section*{Data Availability Statement}
The machine-learning codes and documentation for training PhysNet,
FCHL, and RKHS+F-based models are available at
\url{https://github.com/MMunibas/PhysNet},
\url{https://github.com/qmlcode/qml}, and
\url{https://github.com/MMunibas/RKHS_CH2O} and the reference data can
be downloaded from zenodo \url{https://doi.org/10.5281/zenodo.3923823}.\\

\section*{Acknowledgments}
This work was supported by the Swiss National Science Foundation
grants 200021-117810, 200020-188724, the NCCR MUST, the AFOSR, and the
University of Basel which is gratefully acknowledged (to MM).  We
acknowledge additional support by the Swiss National Science
foundation (No.~NFP 75 Big Data, 200021\_175747, NCCR MARVEL) and from
the European Research Council (ERC-CoG grant QML).  Some calculations
were performed at sciCORE (http://scicore.unibas.ch/), the scientific
computing core facility at University of Basel.

\bibliography{references}
\end{document}

\date{\today}

\begin{table}[h]
\begin{tabular}{l|l|l|l}
  ab initio	&	\textbf{B3LYP}	&	\textbf{MP2}	&	\textbf{CCSD(T)-F12}	\\
\hline
\textbf{CO} [\r{A}]	&	1.2042	&	1.2129	&	1.2069	\\
\textbf{CH}	 [\r{A}]&	1.1206	&	1.1002	&	1.1023	\\
\textbf{HCH} [$^\circ$]&	115.05	&	116.59	&	116.67	\\
\textbf{OCH} [$^\circ$]	&	122.48	&	121.70	&	121.67	
\end{tabular}
\caption{{\it Ab initio} optimized H$_2$CO bond lengths and angles.}
\label{sitab:geom_abinitio}
\end{table}

\begin{table}[h]
\begin{tabular}{l|l|l|l}
PhysNet & \textbf{B3LYP} & \textbf{MP2} & \textbf{CCSD(T)-F12}
\\\hline \textbf{CO} [\r{A}] & 1.2042 & 1.2129 & 1.2069 \\ \textbf{CH}
         [\r{A}] & 1.1206 & 1.1002 & 1.1023 \\ \textbf{HCH} [$^\circ$]
         & 115.05 & 116.59 & 116.67 \\ \textbf{OCH} [$^\circ$] &
         122.47 & 121.70 & 121.67
\end{tabular}
\caption{H$_2$CO bond lengths and angles optimized using PhysNet.}
\label{sitab:geom_physnet}
\end{table}
\begin{table}[h]
\begin{tabular}{l|l|l|l}
RKHS+F & \textbf{B3LYP} & \textbf{MP2} & \textbf{CCSD(T)-F12} \\\hline
\textbf{CO} [\r{A}] & 1.2042 & 1.2129 & 1.2069 \\ \textbf{CH} [\r{A}]
& 1.1206 & 1.1002 & 1.1023 \\ \textbf{HCH} [$^\circ$] & 115.05 &
116.59& 116.67 \\ \textbf{OCH} [$^\circ$] & 122.47 & 121.70 & 121.67
\end{tabular}
\caption{H$_2$CO bond lengths and angles optimized using RKHS+F.}
\label{sitab:geom_rkhs}
\end{table}

\begin{table}[h!]
\begin{tabular}{l|l|l|l}
FCHL & \textbf{B3LYP} & \textbf{MP2} & \textbf{CCSD(T)-F12} \\\hline
\textbf{CO} [\r{A}] & 1.2042 & 1.2129 & 1.2069 \\ \textbf{CH} [\r{A}]
& 1.1206 & 1.1002 & 1.1023 \\ \textbf{HCH} [$^\circ$] & 115.05 &
116.59 & 116.67 \\ \textbf{OCH} [$^\circ$] & 122.47 & 121.70 & 121.67
\end{tabular}
\caption{H$_2$CO bond lengths and angles optimized using FCHL.}
\label{sitab:geom_fchl}
\end{table}

\begin{figure}[!h]
\centering \includegraphics[width=0.6\textwidth]{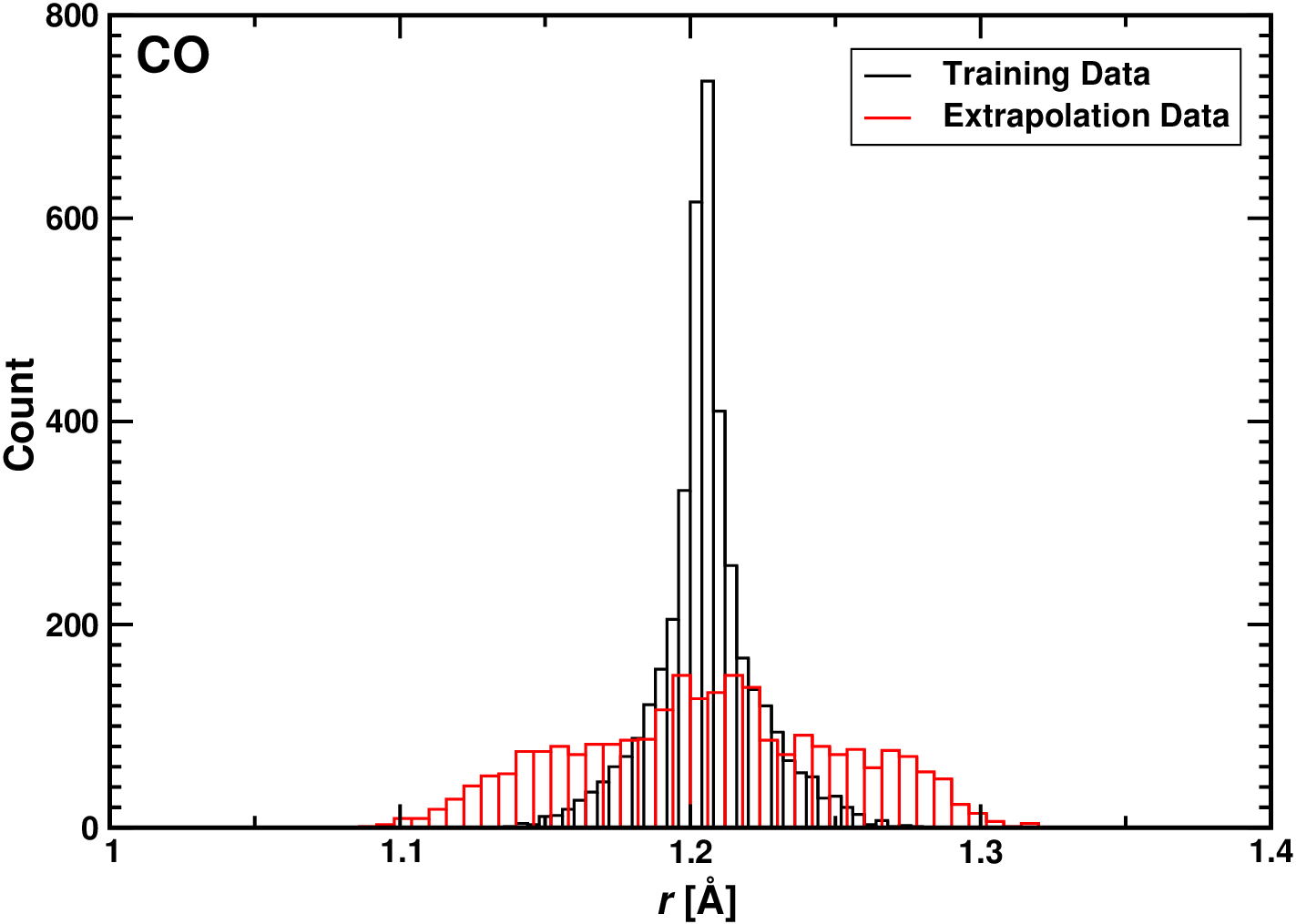}
\caption{Histogram of the CO bond lengths present for Set1
  (black) and Set2 (red).}
\label{sifig:dataset_CO}
\end{figure}

\begin{figure}[!h]
\centering
\includegraphics[width=0.6\textwidth]{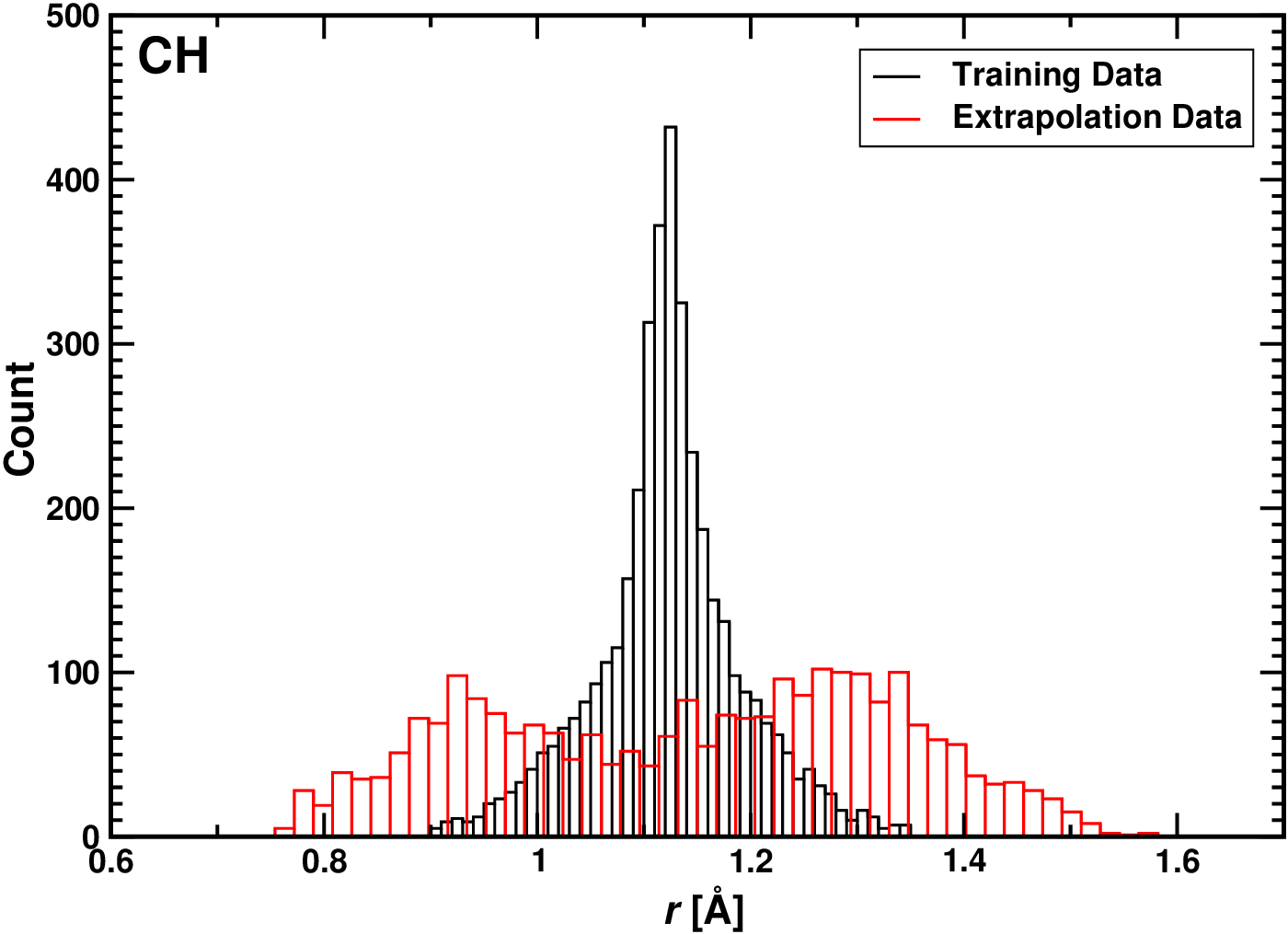}
\caption{Histogram of the CH bond lengths present for Set1
  (black) and Set2 (red).}
\label{sifig:dataset_CH}
\end{figure}

\begin{figure}[!h]
\centering
\includegraphics[width=0.6\textwidth]{oh.eps}
\caption{Histogram of the OH bond lengths present for Set1
  (black) and Set2 (red).}
\label{sifig:dataset_OH}
\end{figure}

\begin{figure}[!h]
\centering
\includegraphics[width=1\textwidth]{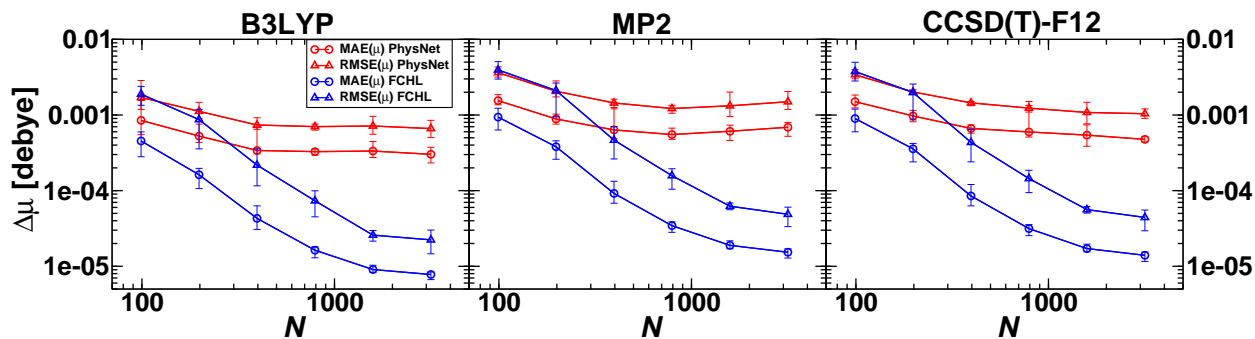}
\caption{Log-log plot of the dipole moment learning curves for PhysNet
  (red) and FCHL (blue). The models are trained with different data
  set sizes (100, 200, 400, 800, 1600, 3200) and on data at different
  levels of theory. The MAE is shown as a circle and the RMSE is shown
  as a triangle. $\Delta \mu$ corresponds to the dipole moment error and the
  error bars indicate the minimum and maximum error. 
  Every data point is an average over 5 models trained
  on the same data set size, but different samples from Set1. A
  PhysNet model trained to convergence of the force and dipole moment
  reaches $\sim 10^{-4}$ debye.}
\label{sifig:dipole_learning_curve}
\end{figure}

\begin{figure}[ht]
\centering
\includegraphics[width=0.9\textwidth]{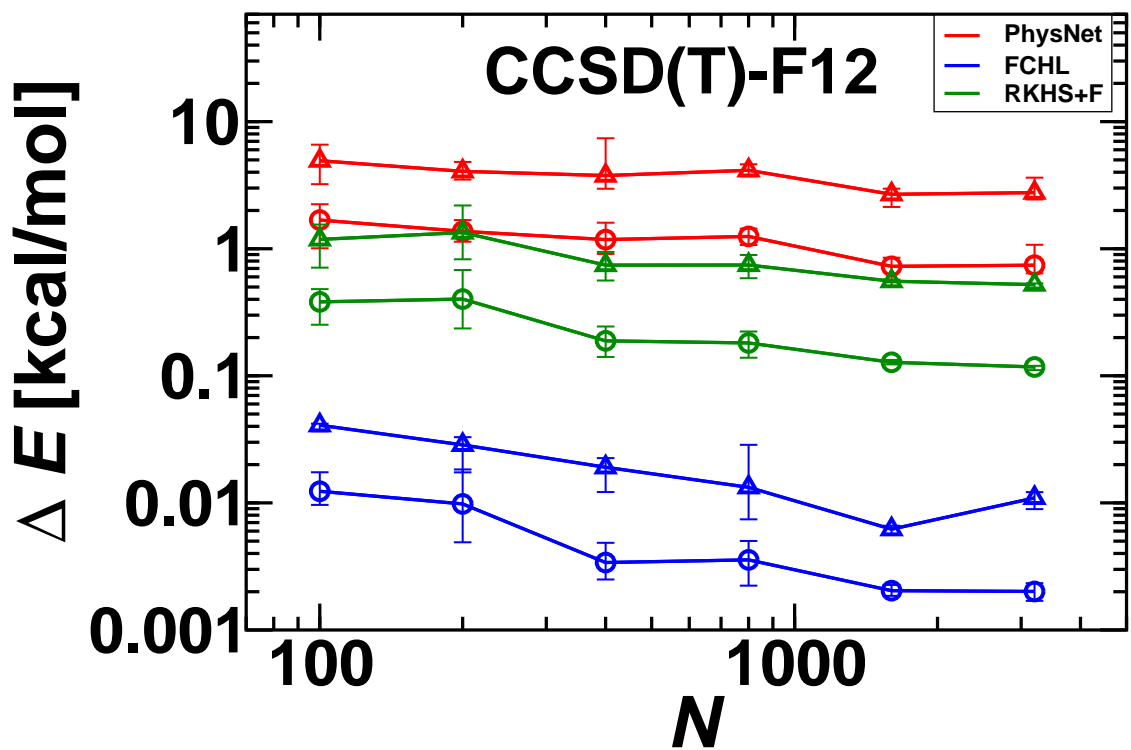}
\caption{Log-log plot of the energy errors for the PhysNet- (red),
  FCHL- (blue) and RKHS+F- (green) based models tested on Set2. Only
  little improvement is found for larger data set sizes. All models
  are trained on the same reference data with different data set sizes
  (100, 200, 400, 800, 1600, 3200) and on data calculated at the
  CCSD(T)-F12 level of theory. The MAE is shown as a circle and the
  RMSE is shown as a triangle. $\Delta E$ corresponds to the energy
  error and the error bars indicate the minimum and maximum
  error. Every data point is an average over 5 models trained
  independently on the same data set size, but different samples from
  the full data set.}
\label{sifig:extrapolation_learning_curve}
\end{figure}

\clearpage

\begin{table}[h]
\centering
\begin{tabular}{c|c|c|c|c|c}
\textbf{mode} & \textbf{G09/PhysNet H} &\textbf{MP2 H} & \textbf{G09/PhysNet AH} & \textbf{MP2 AH} & \textbf{Exp\cite{herndon2005determination}}\\\hline
$\nu_1$ & 2972.1	&	2973.4	&	2861.8	&	2826.8	&	2782.0	\\
$\nu_2$ & 1752.5	&	1753.0	&	1714.5	&	1721.0	&	1746.0	\\
$\nu_3$ & 1539.7	&	1540.1	&	1507.6	&	1508.0	&	1500.0	\\
$\nu_4$ & 1196.4	&	1196.9	&	1188.7	&	1180.2	&	1167.0 \\
$\nu_5$ & 3045.5	&	3047.5	&	2888.3	&	2862.7	&	2843.0 \\
$\nu_6$ & 1266.9	&	1266.9	&	1251.1	&	1246.7	&	1249.0 \\
\hline
RMSE    &  1.03  &  & 18.32 & 23.32 &    
\end{tabular}
\caption{Harmonic and anharmonic frequencies in cm$^{-1}$ for the
  optimized H$_2$CO structure compared with those from
  experiment\cite{herndon2005determination}. The harmonic (G09/PhysNet
  H) and anharmonic (G09/PhysNet AH) frequencies are calculated with
  Gaussian 09 using the PhysNet PES trained on 3200 MP2 energies,
  forces and dipole moments. They are compared to their reference ab
  initio values and center frequencies from experiment
  (Exp). The RMSEs between \textit{ab initio} MP2 frequencies and
  PhysNet predictions (H and AH) and between experiment and MP2 AH frequencies
  are shown in the last row.}\label{sitab:anh_freq_mp2}
\end{table}

\clearpage

\begin{table}[h]
\scalebox{0.82}{
\begin{tabular}{c|c|c|c|c|c|c|c||c}
[kcal/mol] & PhysNet$_{\rm B3LYP}$   & TL(1,1)   & TL(9,1)   & TL(22,3)  & TL(45,5)  & TL(90,10) & TL(180,20) & CCSD(T)-F12(3200) \\\hline
MAE(E)	&	5.011	&	0.148	&	0.045	&	0.021	&	0.011	&	0.005	&	0.004	&	0.000	\\
RMSE(E)	&	5.107	&	0.246	&	0.055	&	0.034	&	0.043	&	0.011	&	0.006	&	0.001	\\
MAE(F)	&	5.198	&	1.391	&	0.167	&	0.105	&	0.128	&	0.055	&	0.040	&	0.007	\\
RMSE(F)	&	7.291	&	2.194	&	0.414	&	0.341	&	0.437	&	0.183	&	0.090	&	0.013	
\end{tabular}}
\caption{Transfer learning from B3LYP to CCSD(T)-F12. The MAEs and
  RMSEs for energies and forces are
  reported. PhysNet$_{\textrm{B3LYP}}$ corresponds to the PhysNet
  trained on B3LYP data but predicting the CCSD(T)-F12 data (see also
  black symbols in Figure~7). TL$(N_{\rm
    train}^{\rm TL},N_{\rm valid}^{\rm TL})$ correspond to the B3LYP
  model transfer learned with $N_{\rm tot}^{\rm TL}= N_{\rm
    train}^{\rm TL} + N_{\rm valid}^{\rm TL}$ data points. The
  CCSD(T)-F12(3200) column corresponds to the performance of PhysNet
  trained and tested on CCSD(T)-F12 data. The energies are in kcal/mol
  and the forces in kcal/mol/\r{A}. The TL models are tested on the
  remaining geometries of the CCSD(T)-F12 data set (i.e. the TL(1,1)
  model is evaluated on 3999, the TL(9,1) on 3991 structures and
  similar for the other models).}
\label{sitab:lc_errors}
\end{table}
\vspace{2cm}

\begin{figure}[h!]
\centering \includegraphics[width=0.8\textwidth]{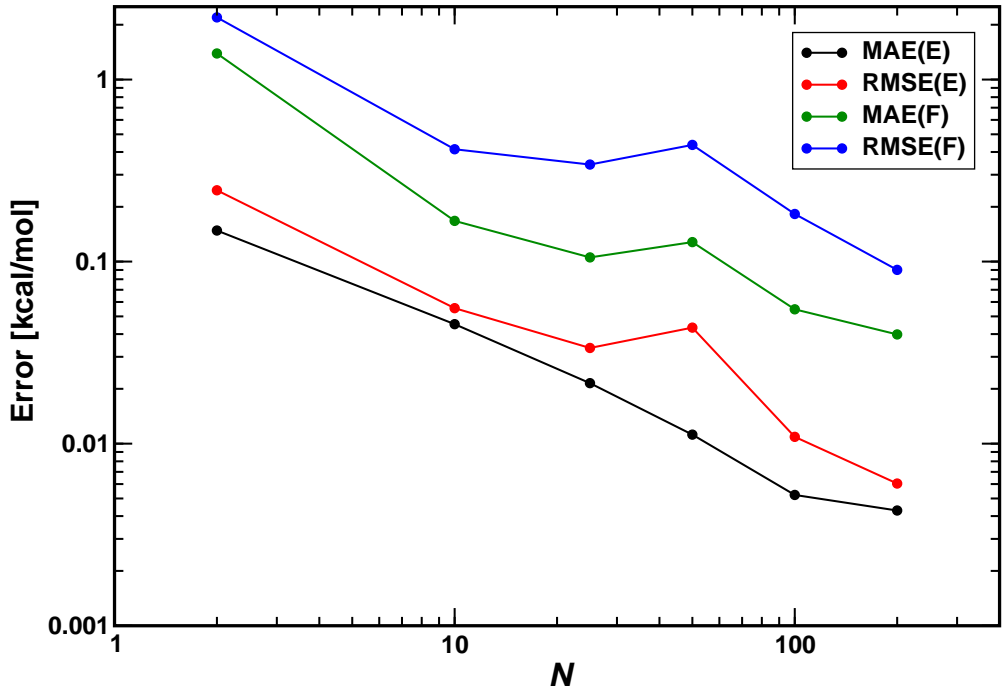}
\caption{Log-log plot of the energy and force learning curves for
  different transfer learned models.  The models are trained with
  different data set sizes $N_{\rm tot}^{\rm TL} = (2, 10, 25, 50,
  100, 200)$.}
\label{sifig:lc_TL}
\end{figure}

\bibliography{references}